\documentclass[12pt, draftclsnofoot, onecolumn]{IEEEtran}

\usepackage{verbatim}
\usepackage[dvips]{graphicx}
\usepackage[cmex10]{amsmath}
\usepackage{amssymb,amsfonts}
\usepackage{multirow}
\usepackage{theorem}
\def\myproof{\noindent{\underline{\textbf{Proof:}}}}

\ifdefined\myproof
\else
\def\myproof{\proof}

\fi

\newtheorem{proof}{Proof}

\newcommand\MYhyperrefoptions
{
	bookmarks=false, 
	bookmarksnumbered=true,
	pdfpagemode={UseOutlines},
	plainpages=false,
	pdfpagelabels=true,
	colorlinks=true,
	linkcolor={black},
	citecolor={black},
	urlcolor={black},
	pdftitle={}, 
	pdfcreator={LaTex},
}

\usepackage{cite}
\usepackage[\MYhyperrefoptions,breaklinks=true,]{hyperref}
\usepackage{breakurl}

\usepackage[tight,footnotesize]{subfigure}

\usepackage{algorithm}
\usepackage{multirow}
\usepackage{algorithmic}

\usepackage{subfigure}

\usepackage[justification=centering]{caption}
\usepackage{bbm}

\usepackage{stfloats}

\def\BibTeX{{\rm B\kern-.05em{\sc i\kern-.025em b}\kern-.08em
		T\kern-.1667em\lower.7ex\hbox{E}\kern-.125emX}}

\begin{document}
\title{Hierarchical Reinforcement Learning for Relay Selection and Power Optimization in Two-Hop Cooperative Relay Network}

\author{Yuanzhe~Geng,
        Erwu~Liu,~\IEEEmembership{Senior~Member,~IEEE,}
        Rui~Wang,~\IEEEmembership{Senior~Member,~IEEE,}
        and~Yiming~Liu,~\IEEEmembership{Student~Member,~IEEE}
\thanks{This work is supported in part by the grants from the National Science Foundation of China (No. 61571330, No. 61771345), Shanghai Integrated Military and Civilian Development Fund (No. JMRH-2018-1075), and Science and Technology Commission of Shanghai Municipality (No. 19511102002). Corresponding author: Erwu Liu.}
\thanks{Yuanzhe Geng, Erwu Liu, and Yiming Liu are with the College of Electronics and Information Engineering, Tongji University, Shanghai 201804, China, E-mail: yuanzhegeng@tongji.edu.cn, erwu.liu@ieee.org, ymliu\_970131@tongji.edu.cn.}
\thanks{Rui Wang is with the College of Electronics and Information Engineering and
 Shanghai Institute of Intelligent Science and Technology, Tongji University,
 Shanghai 201804, China, E-mail: ruiwang@tongji.edu.cn.}
}

\maketitle
\renewcommand\thepage{}

\begin{abstract}
Cooperative communication is an effective approach to improve spectrum utilization. In order to reduce outage probability of communication system, most studies propose various schemes for relay selection and power allocation, which are based on the assumption of channel state information (CSI). However, it is difficult to get an accurate CSI in practice.
In this paper, we study the outage probability minimizing problem subjected to a total transmission power constraint in a two-hop cooperative relay network. We use reinforcement learning (RL) methods to learn strategies for relay selection and power allocation, which do not need any prior knowledge of CSI but simply rely on the interaction with communication environment. It is noted that conventional RL methods, including most deep reinforcement learning (DRL) methods, cannot perform well when the search space is too large.
Therefore, we first propose a DRL framework with an outage-based reward function, which is then used as a baseline. Then, we further propose a hierarchical reinforcement learning (HRL) framework and training algorithm. A key difference from other RL-based methods in existing literatures is that, our proposed HRL approach decomposes relay selection and power allocation into two hierarchical optimization objectives, which are trained in different levels. With the simplification of search space, the HRL approach can solve the problem of sparse reward, while the conventional RL method fails.
Simulation results reveal that compared with traditional DRL method, the HRL training algorithm can reach convergence 30 training iterations earlier and reduce the outage probability by $5\%$ in two-hop relay network with the same outage threshold.

\end{abstract}


\section{Introduction}\label{sect_intro}
The rapid development of communication technology makes wireless spectrum resources very tight \cite{8275026,7886242,Liu2012:Energy}.
Therefore, in recent years, cooperative communication has been paid much attention, for it helps improving spectrum utilization and system throughput in multi-user scenario.

Cooperative communication systems usually use outage probability as a metric, to measure the Quality-of-Service (QoS) and the robustness of the system.
An outage occurs when the received signal-to-noise ratio (SNR) falls below a certain threshold \cite{8435942, 2006.00664}.
In order to minimize the outage probability of the cooperative relay network, there are usually two intuitive approaches, that is, optimize relay selection scheme or power allocation scheme.

\textbf{Relay Selection Schemes:}
For a scenario with multi-relay, it is usually possible to select multiple relays to coordinate data transmission and assign orthogonal channels to avoid interference.
Jedrzejczak \textit{et al.} \cite{7438886} studied the relay selection problem by calculating of harmonic mean of channel gains.
Islam \textit{et al.} \cite{7778750} demonstrated the influence of network coverage capability on relay selection.
Das and Mehta \cite{7105886} proposed an approach for relay selection by analyzing the outage probability.
The draw back of employing too many relays is that, it may lead to extensive time consumption and frequency resource waste when forwarding signal.
To solve this issue, Bletsas \textit{et al.} \cite{1603719} proposed an opportunistic relay section scheme by only choosing the best relay according to channel state, which can obtain the full set gain.
However, these methods all assume an exact channel state information (CSI), which is not practical because of the existence of inevitable noise.

\textbf{Power Allocation Schemes:}
Based on the given relay, reasonable power allocation can further improve the received SNR and reduce the outage probability.
Given partial channel information, Wang and Chen \cite{7313006} derived a closed-form formula for optimal power allocation based on maximizing a tight capacity lower bound.
In \cite{5722051} and \cite{6108302}, the authors considered optimal power allocation scheme in different situations of convention relay and opportunistic relay.
Tabataba \textit{et al.} \cite{5618893} deduced the expression of system outage probability under the condition of high SNR, and studied power allocation using AF protocol.
However, these researches still require prior knowledge of the channel, which can not be further applied to other situations.

Recently, some researchers have successfully applied reinforcement learning (RL) methods to cooperative communication.
RL is one of the three paradigms of machine learning, which can achieve high-precision function fitting through powerful computing capability.
Unlike traditional methods, RL methods do not need prior knowledge of the environment, that is, we do not have to add any assumptions to the learning process.

In RL approaches, the source node is empowered with the learning ability to determine the optimal relay or power allocation for the current moment, based on previous observation of system state and rewards.
Shams \textit{et al.} \cite{6954557} employed Q-learning algorithm to solve the power control problem,
and Wang \textit{et al.} \cite{9072416} proposed a Q-learning based relay selection scheme in relay-aided communication scenarios.
The drawback of these studies is obvious, as they can be only suitable for simple problems with a low dimension.
Su \textit{et al.} \cite{8750861} proposed a deep Q network (DQN) based relay selection scheme with detailed mutual information (MI) as reward, but did not take power allocation into consideration.
In order to solve the joint optimization problem, Su \textit{et al.} \cite{9137340} employed convex optimization and DQN to deal with relay selection and power allocation, respectively.
However, these RL methods and frameworks can only be used in some simple environments, and do not work well in the high-dimensional search space.

In this paper, we propose a hierarchical reinforcement learning (HRL) approach for relay selection and power allocation, to minimize the outage probability of the two-hop cooperative communication system.
Unlike traditional optimization methods, our method can learn behavior policy without assuming any prior knowledge of channel state.
It is also different from existing RL-based methods that, we design an outage-based reward function, which uses a binary signal to represent success or failure of communication.
Furthermore, we propose a novel hierarchical framework to reduce searching space and improve learning efficiency.
Specifically, the contributions of this paper can be summarized as follows.
\begin{itemize}
    \item
    In our two-hop cooperative communication model, we transform the traditional outage probability optimization problem into a statistical problem, so that the RL method can be used to solve the problem.
    By employing RL methods, we no longer need to add any assumptions to channel distributions, and only rely on interaction with communication environment.

    \item
    We propose an outage-based reward function.
    Compared with other existing RL methods, our method needs less information fed back from the environment.
    Rewards are only determined by binary signals of success or failure, and do not include other concrete representations of information.
    It is practical because other additional feedback may be not available in certain situations.

    \item
    We further design an HRL framework with two levels for cooperative relay network, where relay selection and power allocation are disassembled into two optimization objectives.
    Traditional deep reinforcement learning (DRL) methods considers relay selection and power allocation together, which leads to a more complex action space and may affected the learning performance.
    By decomposing different optimization objectives into different levels, complex action space is therefore simplified in our HRL framework.
\end{itemize}

The rest of this paper is organized as follows.
Section \ref{sect preli} introduces the preliminaries of DRL.
Section \ref{sect model} analyzes our system model and Section \ref{sect problem} formulates the outage minimization problem.
Section \ref{sect DRL} describes our outage-based method using DQN framework.
Section \ref{sect HRL} describes our proposed HRL framework and learning algorithm, and presents our pre-training algorithm in detail.
Section \ref{sect Result} presents simulation results.
Finally, Section \ref{sect Conclusion} concludes this paper and outlines future works.

\section{Preliminaries}\label{sect preli}
In the field of cooperative communication, recent studies have proposed several machine learning methods for relay selection or power allocation.
Traditional communication methods make assumptions about CSI, while these methods use data for learning and then making channel predicting.

As one of the three paradigms of machine learning, RL is an emerging tool to solve decision-making problems such as resource management in communication \cite{8931561,8761525,8771176}.
RL methods use an agent, which can be regarded as an intelligent robot, to interact with the environment.
The agent in RL methods has no access to prior knowledge of the environment, but can only get familiar with the environment through an interactive process called Markov Decision Process (MDP).
In order to minimize outage probability, the agent will repeatedly interact with the communication environment, and choose a suitable relay and allocate transmission power according to current system state.
In addition, it continuously adjusts its behavior policy according to the feedback from the communication environment.
In this section, we introduce these preliminaries of reinforcement learning.

\subsection{Markov Decision Process}\label{subsect_preliA}
An MDP consists of an environment $\mathcal{E}$, a state space $\mathcal{S}$, an action space $\mathcal{A}$, and a reward function $\mathcal{S}\times\mathcal{A}\to\mathcal{R}$. At each discrete time step $t$, the agent observes the current state $s_t\in\mathcal{S}$, and selects an action $a_t\in\mathcal{A}$ according to a policy $\pi$: $\mathcal{S}\to \mathcal{P}(\mathcal{A})$, which maps states to a probability distribution over actions. After executing action $a_t$, the agent receives a scalar reward $r(s_t,a_t)$ from the environment $\mathcal{E}$ and observes the next state $s_{t+1}$ according to the transition probability $p(s_{t+1}|s_t,a_t)$. This process will continue until a terminal state is reached.

The goal of the agent is to find the optimal policy to maximize the expected long-term discounted reward, \emph{i.e.}, maximize the expected accumulated return $R_t=\sum\nolimits_{i=t}^T \gamma^{i-t}r(s_i,a_i)$ from each state $s_t$, where $T$ denotes the total step, and $\gamma\in[0,1]$ denotes the discount factor that trades off the importance of immediate and future rewards.

Action-value function is usually used to describe the expected return after selecting action $a_t$ in state $s_t$ according to policy $\pi$.
\begin{equation}\label{eq2_1} Q^{\pi}(s_t,a_t)=\mathbb{E}_{s_t\sim\mathcal{E},a_t\sim\pi(s_t)}(R_t|s_t,a_t). \end{equation}
And we can obtain the preceding action-value function via recursive relationship known as Bellman function.
\begin{equation}\label{eq2_2} \begin{aligned}
Q^{\pi}(s_t,a_t)=\mathbb{E}_{s_{t+1}\sim\mathcal{E}}\Big[r(s_t,a_t)+\gamma\mathbb{E}_{a_{t+1}\sim\pi(s_{t+1})}\big[Q^{\pi}(s_{t+1},a_{t+1})\big]\Big].
\end{aligned}\end{equation}
Moreover, the optimal action-value function $Q^{\ast}(s_t,a_t)=\max_{\pi\in\Pi}Q^{\pi}(s,a)$ gives the maximum action value under state $s$ and action $a$, and it also obeys Bellman function.
\begin{equation}\label{eq2_3} Q^{\ast}(s,a)=\mathbb{E}_{s^{\prime}\sim\mathcal{E}}[r+\gamma\max\limits_{a^{\prime}}Q^{\pi}(s^{\prime},a^{\prime})]. \end{equation}

\subsection{Reinforcement Learning}\label{subsect_preliB}
In practice, we usually do not know the underlying state transition probability, \textit{i.e.}, in a model-free situation. It requires the agent to interact with the environment and learn from the feedback, constantly adjust its behavior to maximize the expected reward.

In reinforcement learning, temporal difference (TD) methods \cite{RL:intro2} are proposed via combining Monte Carlo methods and dynamic programming methods, which enable the agent to learn directly from raw experience. Therefore, we have the following well-known Q-learning algorithm.
\begin{equation}\label{eq2_5} Q(s_t,a_t)\leftarrow Q(s_t,a_t)+\alpha\delta_t \end{equation}
with
\begin{equation}\label{eq2_4} \delta_t=r(s_t,a_t)+\gamma\max\limits_{a_{t+1}\in\mathcal{A}}Q(s_{t+1},a_{t+1})-Q(s_t,a_t), \end{equation}
where $\delta_{t}$ denotes TD error and $\alpha\in[0,1]$ denotes learning rate.

Through continuous iterative updating, the Q value of different actions selected in each state finally tends to be stable, which can then provide a policy for the subsequent action selection.

\section{System Model}\label{sect model}
\begin{figure}[ht]
    \centering
    \includegraphics[scale=0.75]{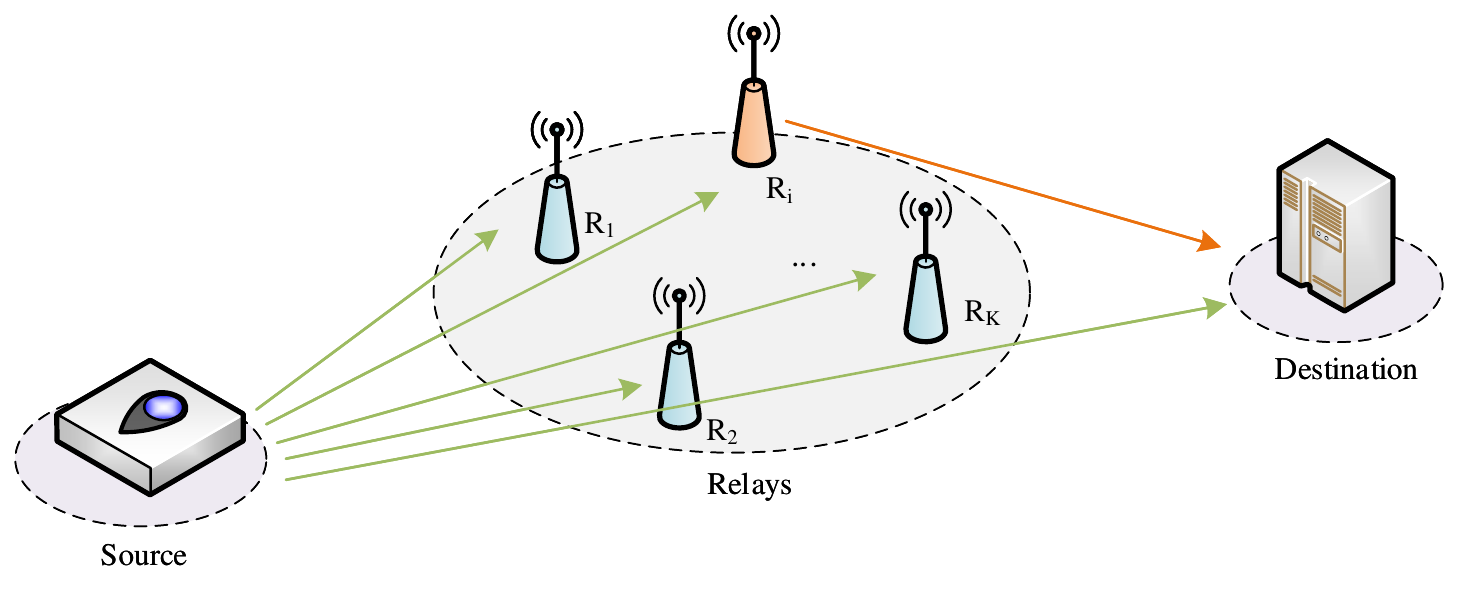}
    \caption{Relay network model.}
    \label{model}
  \end{figure}

Consider a wireless network where exists an $N_S$-antenna source $S$, an $N_D$-antenna destination $D$, and a group of single-antenna relays $R=\{R_1,R_2,\dots,R_K\}$, as shown in Fig. \ref{model}.
We assume that the source is far from the destination, and the help of relay nodes is needed.
Due to the limitation of equipment of relays, we consider a half-duplex signaling mode where the communication from $S$ to $D$ via the selected relay $R_i$ will take two time slots.
In the first time slot, $S$ broadcasts its signal, and all the other nodes, include the destination, listen to this transmission.
In the second time slot, the selected relay forwards the decoded signal to destination.

Depending on how the cooperative relay processes the received signal, the relay mode can be mainly divided into amplify-and-forward (AF) and decode-and-forward (DF).
Next, we will analyze the MI obtained by using AF protocol and DF protocol, respectively.

\subsection{Amplify-and-Forward Relaying}\label{subsect_AF_Mode}
In the case that all relays can only scale the received signal and send it to the destination, we employ AF protocol to realize cooperative communication. In the first phase, the received signal at $R_i$ can be written as
\begin{equation}\label{eq3_1} y_{si}(t)=\sqrt{P_s}\boldsymbol h_{si}^{\dagger}(t) \boldsymbol x(t)+n_i(t), \end{equation}
where $P_s \in [0,P_{max}]$ is the transmission power at source with $P_{max}$ being the maximum value, $\boldsymbol x(t)$ is a $N_s \times 1$ data symbol vector and $\|\boldsymbol x(t)\|=1$,
$\boldsymbol h_{si}(t)=[h_{si}^1(t),\dots,h_{si}^{N_S}(t)]^T$ is an $N_S\times 1$ channel vector between source and relay,
and each element is a complex Gaussian random variable with zero mean and variance $\sigma_{si}^2$, $n_i(t)\sim\mathcal{CN}(0,\sigma_n^2)$ is the complex Gaussian noise at relay. Similarly, we have the received signal at $D$ which can be written as
\begin{equation}\label{eq3_2} \boldsymbol y_{sd}(t)=\sqrt{P_s}\boldsymbol h_{sd}^{\dagger}(t)\boldsymbol x(t)+\boldsymbol n_d(t), \end{equation}
where $\boldsymbol h_{sd}(t)$ denotes a $N_S \times N_D$ channel matrix, and $\boldsymbol n_d(t)\sim \mathcal{CN}(\boldsymbol 0,\sigma_n^2\boldsymbol I_{N_D})$ denotes the complex Gaussian noise at destination, where $\boldsymbol I$ is identity matrix.

In the second time slot, the selected relay amplifies the signal and transmits it to the destination.
The received signal at the destination from the relay can be written as
\begin{equation}\label{eq3_3} \begin{aligned}
&\boldsymbol y_{id,AF}(t)=\sqrt{P_r}\boldsymbol h_{id}^{\dagger}(t)\beta y_{si}(t)+\boldsymbol n_d(t), \\
\end{aligned}\end{equation}
where $P_r\in [0,P_{max}]$ is the transmission power at relay with $P_{max}$ being the maximum value,
$\boldsymbol h_{id}(t)=[h_{id}^1(t),\dots,h_{id}^{N_D}(t)]$ is a $1 \times N_D$ channel vector between relay and destination,
and similarly, each element is a complex Gaussian random variable with zero mean and variance $\sigma_{id}^2$.
$\beta$ is the amplification factor, which can be written as follows \cite{5710995, 1427716}.
\begin{equation}\label{eq3_4} \beta =\sqrt{\frac{1}{P_s\|\boldsymbol h_{si}\|^2+\sigma^2_n}}. \end{equation}

The destination combines the data from the source and the relay using maximal ratio combining (MRC), and after some manipulations according to \cite{1350931,4107949}, we have the following final end-to-end SNR.
\begin{equation}\label{eq3_5} \varphi_{z} =\frac{\varphi_{si}\varphi_{id}}{\varphi_{si}+\varphi_{id}+1}, \end{equation}
where $\varphi_{si}=P_s{\|\boldsymbol h_{si}\|}^2/\sigma_n^2$ and $\varphi_{id}=P_r{\|\boldsymbol h_{id}\|}^2/\sigma_n^2$.

Similarly, we can obtain the SNR of direct transmission from source to destination, which can be represented as $\varphi_{sd}=P_s{\|\boldsymbol h_{sd}\|}^2/\sigma_n^2$. Then we have the MI between the source and the destination using AF protocol.
\begin{equation}\label{eq3_7} \begin{aligned}
I_{AF}=\frac{1}{2}\log_2(1+\varphi_{AF})=\frac{1}{2}\log_2(1+\varphi_{sd}+\frac{\varphi_{si}\varphi_{id}}{\varphi_{si}+\varphi_{id}+1}).
\end{aligned}\end{equation}


\subsection{Decode-and-Forward Relaying}\label{subsect_DF_Mode}
Assume that all relays are able to decode the signal from the source, and then re-encode and transmit the signal to the destination. The first time slot in DF mode is the same as that in AF mode, and (\ref{eq3_1}) and (\ref{eq3_2}) have given the received signal at $R_i$ and $D$ in this time slot.

In the second time slot, different from that in AF mode, the selected relay decodes and forwards the signal to the destination, and the received signal at the destination from the relay can be written as
\begin{equation}\label{eq3_8} \boldsymbol y_{id,DF}(t)=\sqrt{P_r}\boldsymbol {h_{id}}^{\dagger}(t) y_{si}(t)+\boldsymbol n_d(t), \end{equation}

When employing DF protocol, the relays need to first successfully decode the signal from source on the condition that the MI is above the required transmission rate. Then, the signals received in both two time slots are combined at the destination using MRC \cite{4107949,1603719}, and we have the following instantaneous MI between the source and the destination.
\begin{equation}\label{eq3_9} \begin{aligned}
I_{DF}=\frac{1}{2}\log_2(1+\varphi_{DF})=\frac{1}{2}\log_2(1+\varphi_{sd}+\varphi_{id}).
\end{aligned}\end{equation}

\section{Problem Formulation}\label{sect problem}
Suppose that there is an agent in the communication environment, which has access to channel state in previous time slots.
The agent estimates current channel state based on historical CSI, and accordingly selects relay and allocates transmission power.
Afterwards, it receives a reward from the environment, which indicates whether the communication is successful.

In this section, we model this process as an MDP, where historical channel state is regarded as system state, and relay selection along with power allocation are considered as system action.
Then we describe the variables in our two-hop cooperative communication scenario and formulate our optimization problem.

\subsection{State Space}\label{subsect state}
Full observation of our two-hop communication system consists of the channel states between any two nodes in the previous time slot. Therefore, the state space in current time slot is a union of different wireless channel states, which can be denoted as
\begin{equation}\label{eq4_1} \mathcal{S}_t \triangleq [\boldsymbol h_{si}(t-1), \boldsymbol h_{id}(t-1), \boldsymbol h_{sd}(t-1)], \end{equation}
where the integer $i$ satisfies $i \in [1,K]$.

In order to characterize the temporal correlation between time slots for each channel, we employ the following widely adopted Gaussian Markov block fading autoregressive model \cite{5710995, ref1}.
\begin{equation}\label{eq4_2} \boldsymbol h_{ij}(t)=\rho\boldsymbol h_{ij}(t-1)+\sqrt{1-\rho^2}\boldsymbol e(t), \end{equation}
where $\rho$ denotes the normalized channel correlation coefficient between corresponding elements in $\boldsymbol h(t)$ and $\boldsymbol h(t-1)$, $\boldsymbol e(t)\sim\mathcal{CN}(\boldsymbol 0,\sigma^2\boldsymbol I)$ denotes the error variable and is uncorrelated with $\boldsymbol h_{ij}(t)$. According to Jake's fading spectrum, we have $\rho=J_0(2\pi f_d\tau)$ where $J_0(\cdot)$ denotes zeroth-order Bessel function of the first kind, $f_d$ and $\tau$ denote Doppler frequency and the length of time slot, respectively.

\subsection{Action Space}\label{subsect action}
Full action space includes relay selection $\boldsymbol a^{R}(t)$, source power allocation $\boldsymbol a^{P_s}(t)$, and relay power allocation $\boldsymbol a^{P_r}(t)$.

Considered that the total power is constraint, \textit{i.e.}, $P_{s}+P_{r} \leq P_{max}$, we can assume that the sum of power used by the source and its selected optimal relay is $P_{max}$. So $P_r$ can be directly represented by the difference between $P_{max}$ and $P_s$. Then, we can reduce the number of actions that need to be optimized and derive the following reduced action space.
\begin{equation}\label{eq4_3 action space} \mathcal{A}_t \triangleq [\boldsymbol a^{R}(t), \boldsymbol a^{P_s}(t)]. \end{equation}

The first part of action space is relay selection, which is denoted by
\begin{equation}\label{eq4_3 relay selection} \boldsymbol a^{R}(t)=[a_1^R(t),a_2^R(t),\dots,a_K^R(t)], \end{equation}
where $a_k^R(t)=1$ means relay $R_k$ is selected in time slot $t$ and $a_k^R(t)=0$ otherwise.

The second part of is power allocation for the source node. Similarly, it is denoted by
\begin{equation}\label{eq4_3_power allocation} \boldsymbol a^{P_s}(t)=[a_1^{P_s}(t),a_2^{P_s}(t),\dots,a_{L-1}^{P_s}(t)], \end{equation}
where $P_{max}$ is divided into $L$ power-levels, and $a_l^{P_s}(t)=1$ means the $l$-th power-level is selected for source node transmission in time slot $t$, and $a_l^{P_s}(t)=0$ otherwise.

  \begin{figure*}[t]
    \centering
    \includegraphics[scale=0.8]{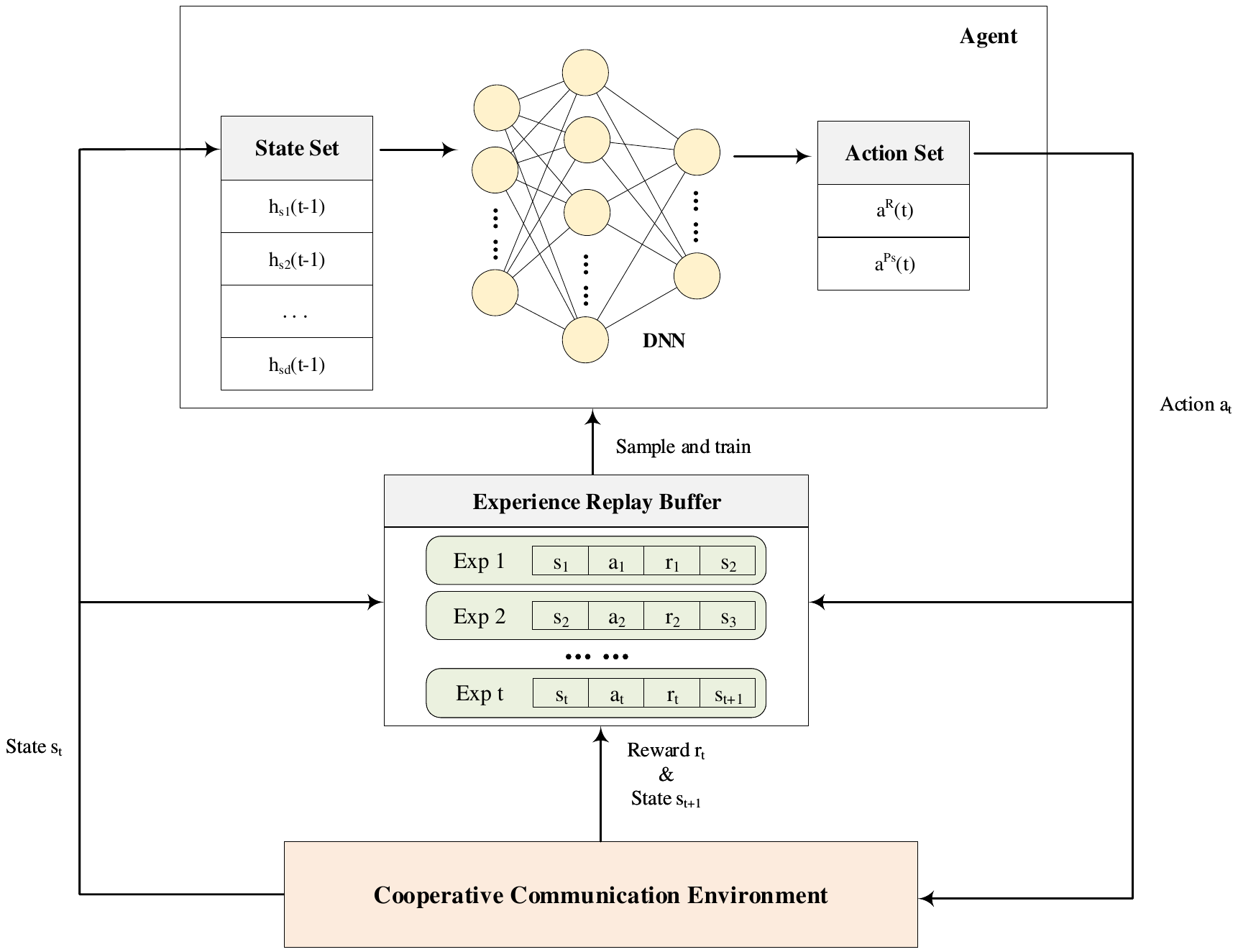}
    \caption{DRL framework for relay selection and power allocation.}
    \label{DQN}
  \end{figure*}

\subsection{Reward and Optimization Problem}\label{subsect reward}
The outage probability minimizing problem that jointly optimizes relay selection and power allocation can be intuitively formulated as
\begin{equation}\label{eq_problem_1}
    \min\limits_{\mathcal{A}_t} \ {\rm \textbf{Prob}}(I<\lambda),
\end{equation}
where the positive scalar $\lambda$ is denoted to be the outage threshold, and $I$ can be calculated according to (\ref{eq3_7}) and (\ref{eq3_9}).

Traditional methods establish a probabilistic model, where the distribution employed to describe channel uncertainty is assumed artificially.
However, we do not rely on underlying channel distributions in this paper, and thus traditional probabilistic analysis methods are not applicable.

In our problem, the agent can only make use of the communication result that denotes success or failure from the cooperative communication environment.
Therefore, we define the following indicator function of event $I<\lambda$, which represents the result after each selection.
\begin{equation}\label{eq4_4_indicator}
    f(\boldsymbol a^{R}, \boldsymbol a^{P_s}; \boldsymbol h) \triangleq \mathbbm{1}_{I<\lambda}=\left\{
        \begin{aligned}
            &1, &if\ I<\lambda\ \ \\
            &0, &otherwise
        \end{aligned}
    \right.
\end{equation}

Consider the fact that, when an indicator function is employed to represent each occurrence of an event, then the expectation of the indicator function can be used to calculate the probability of the original event.
Therefore, we can reformulate the optimization problem for minimizing outage probability of our communication system in the form of statistics.
Then, this problem can be solved using RL methods, which is formulated as follows.
\begin{equation}\label{eq4_4_problem} \begin{aligned}
\min\limits_{\mathcal{A}_t} \ &\mathbb{E} \left[ \frac{1}{T}\sum\limits_{t=1}^T f\big(\boldsymbol a^{R}(t), \boldsymbol a^{P_s}(t);\boldsymbol h(t)\big) \right] \\
s.t.\
&\textbf C_1: \sum_{k=1}^K a_k^R(t)=1, \\
&\textbf C_2: \sum_{l=1}^{L-1} a_l^{P_s}(t)=1, \\
&\textbf C_3: a_k^R(t),a_l^{P_s}(t) \in \{0,1\}.
\end{aligned}\end{equation}

In MDP, the reward is fed back to the agent to evaluate the selected action under current system state.
In this paper, we design an outage-based reward function, which only consists of the communication result.
Note that, the goal of the agent is to find the optimal behavior policy to maximize the expected long-term discounted reward, so our binary reward function is denoted as
\begin{equation}\label{eq4_4_reward}
r_t = 1-f\big(\boldsymbol a^{R}(t), \boldsymbol a^{P_s}(t);\boldsymbol h(t)\big).
\end{equation}

\section{DRL Based Solution}\label{sect DRL}
In reinforcement learning, we often estimate the action-value function by using Bellman function as an iterative update to converge to the optimal.
Unfortunately, traditional RL method which only employs Bellman function has no generalization ability.
Since channel state is uncountable, traditional RL method often fails to make decisions when faced with channel state that has never appeared.
DRL is a combination of deep neural networks (DNN) and traditional RL method, which is proposed to solve generalization problem in large state or action space \cite{Mnih2015, IMPALA}.
In this section, we adopt DQN framework to design a DRL-based solution for relay selection and power allocation.

The DRL framework for relay selection and power allocation is shown in Fig. \ref{DQN}.
Note that the source node has no prior knowledge of the communication system, which means the distributions of wireless channels between any two nodes are all unknown to it.
The agent can only observe the current state from the communication environment and get state $s_t$.
Then a deep neural network is employed as the nonlinear function approximator to deal with the input data, which can be applied to estimate the action-value function for high dimensional state space.
According to (\ref{eq2_3}), we have
\begin{equation}\label{eq4.5_5}
Q^{\pi}(s_t, a_t; \theta)=\mathbb{E}\Big[r_{e,t}+\gamma_l\max\limits_{a_{t+1}}Q^{\pi}(s_{t+1}, a_{t+1}; \theta)\Big].
\end{equation}

After calculation, the agent chooses the best action which can induce the maximal value of $Q^{\pi}(s_t, a_t; \theta)$, and then the environment will give the corresponding reward and update system state.
So far, we have obtained a complete experience tuple $e_t=(s_t,a_t,r_{t},s_{t+1})$, which will be stored in experience replay buffer $\mathcal{B}=\{e_1,e_2,\dots,e_t\}$.
When training, a batch of experience will be sampled and used to optimize a set of loss functions below.
\begin{equation}\label{eq4.5_6} L_i(\theta_i)=\mathbb{E}_{e_t\sim\mathcal{B}}\Big[\big(y_i-Q(s_t, a_t; \theta_i)\big)^2\Big], \end{equation}
with
\begin{equation}\label{eq4.5_7} y_i=r + \gamma_l \max\limits_{a_{t+1}}Q^{\pi}(s_{t+1}, a_{t+1}; \theta^{-}), \end{equation}
where $\theta^{-}$ is parameters from previous iteration in a separate target network which are held fixed when optimizing, and will be replaced by $\theta_{i-1}$ from the evaluate network after a period of time.

Then differentiate operation on these loss functions will be carried out, and we can yield the following expression.
\begin{equation}\label{eq4.5_8} \begin{aligned}
\nabla_{\theta_i}L_i(\theta_i)=\mathbb{E}\Big[\big(y_i-Q(s_t,a_t;\theta_i)\big) \nabla_{\theta_i}Q(s_t,a_t;\theta_i)\Big].
\end{aligned}\end{equation}

The following standard non-centered RMSProp optimization algorithm \cite{A3C,lecture} is then adopted to minimize the loss function and update parameters in Q network .
\begin{equation}\label{eq4.5_9} \theta\leftarrow\theta-\eta\frac{\Delta\theta}{\sqrt{\upsilon+\epsilon}}, \end{equation}
with
\begin{equation}\label{eq4.5_10} \upsilon=\kappa \upsilon +(1-\kappa)\Delta\theta^2, \end{equation}
where $\kappa$ is a momentum and $\Delta\theta$ is the accumulated gradients.

Note that, this is a model-free approach, as the agent using state and reward sampled from the environment rather than estimating transition probability. And this is an offline policy, because an epsilon greedy method will be employed as the behavior policy.
For a two-hop cooperative relay network, Algorithm \ref{algoDQN} employs DQN framework to make dynamic relay selection and power allocation.
In the Evaluation part of this paper, we will test the algorithm and use it for comparison.
Pseudocode of the algorithm can be found in Algorithm \ref{algoDQN}.

\begin{algorithm}[htb]
    \caption{DRL Based Relay Selection and Power Allocation}
    \label{algoDQN}
    \begin{algorithmic}[1]
        \STATE Initialize experience replay buffer $\mathcal{B}$.
        \STATE Initialize Q network with random weights $\theta$.
        \STATE Initialize target Q network with $\theta^{-}=\theta$.
        \FOR{episode $u=1,2,\dots,u_{max}$}
            \STATE Initialize the environment, get state $s_1$.
            \FOR{time slot $t=1,2,\dots,t_{max}$}
                \STATE Choose action $a_t$ using epsilon-greedy method with a fix parameter $\epsilon$.
                \STATE Execute action $a_t$, and observe reward $r_{t}$ and next state $s_{t+1}$.
                \STATE Collect and save the tuple $e_t$ in $\mathcal{B}$.
                \STATE Sample a batch of transitions $(s_j,a_j,r_{j},s_{j+1})$ from $\mathcal{B}$.
                \IF {episode terminates at time slot $j+1$}
                    \STATE Calculate $y_j$ according (\ref{eq4.5_7});
                \ELSE
                    \STATE Set $y_j=r_{j}$.
                \ENDIF
                \STATE Perform gradient descent and update Q-network according to (\ref{eq4.5_8}).
                \STATE Update current state, and every $C$ steps reset $\theta^{-}=\theta$.
            \ENDFOR
        \ENDFOR
    \end{algorithmic}
\end{algorithm}

\section{HRL Based Solution}\label{sect HRL}
Traditional DRL-based approaches put all variables together in its action, which results in a complex search space.
HRL is a recent technology based on DRL, which has developed rapidly in recent years and is considered a promising method to solve problems with sparse rewards in complex environment.
HRL enables more efficient exploration of the environment by abstracting complex tasks into different levels \cite{HDQN, HIRO, 8629360}.
In this section, we propose a novel two-level HRL framework for cooperative communication, to learn relay selection policy and power allocation policy in different levels.
  \begin{figure*}[ht]
    \centering
    \includegraphics[scale=0.85]{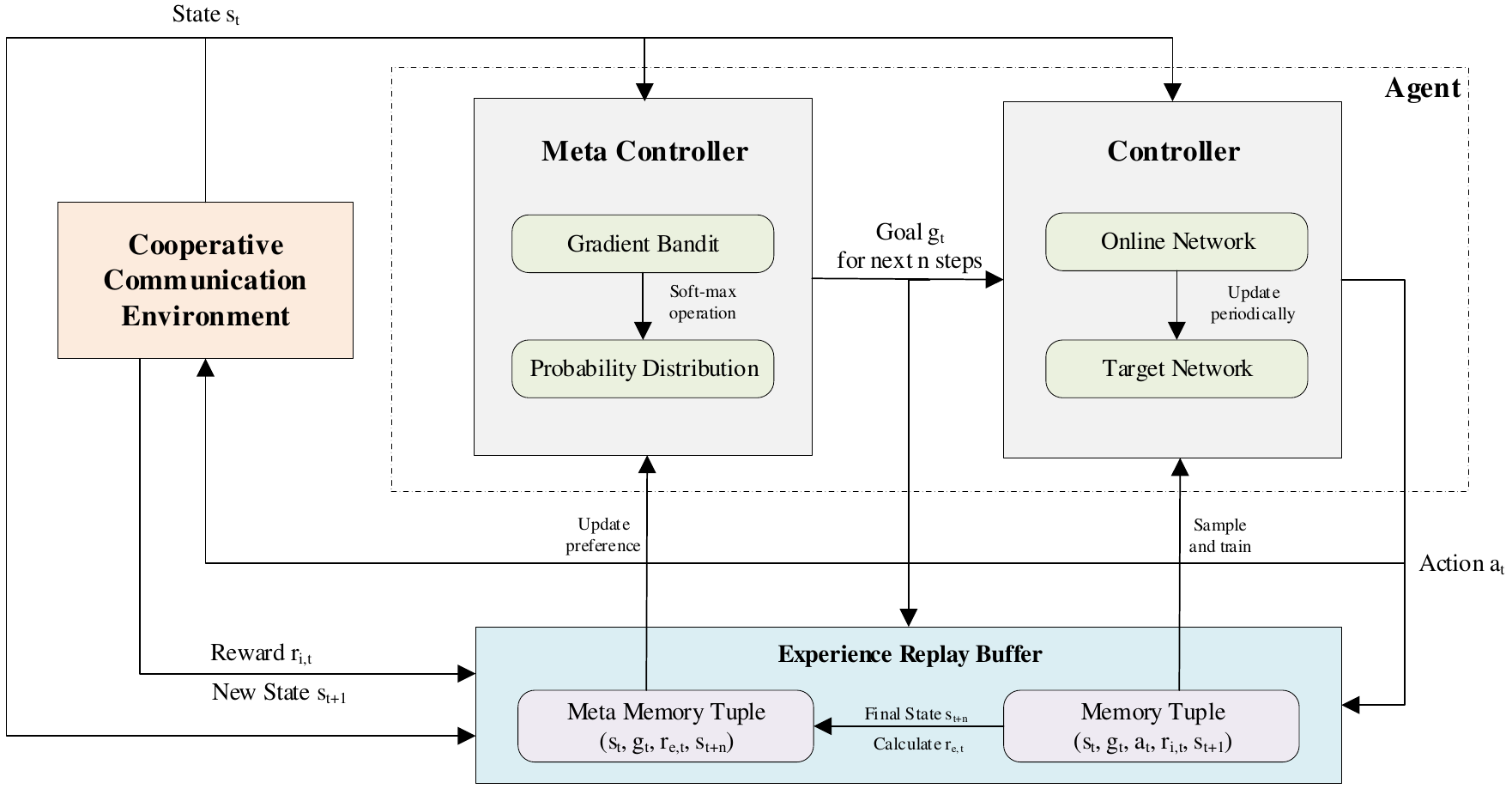}
    \caption{HRL framework for decomposing relay selection and power allocation into different levels.}
    \label{Framework}
  \end{figure*}

\subsection{Proposed Framework}\label{subsect_HRLA}

As shown in Fig. \ref{Framework}, the communication agent has two levels. In the higher level, the meta-controller receives observation of state from the external communication environment, and outputs a goal $g_t$. The controller in the lower level is supervised with the goals that are learned and proposed by meta-controller, which observes the state of the external communication environment and selects an action $a_t$. Note that the meta-controller gives a goal every $n$ steps, and the goal will remain until the low-level controller reaches the terminal. We employ a standard experience replay buffer, and it is worth to note that, experience tuples $(s_t,g_t,r_{e,t},s_{t+n})$ for meta controller and $(s_t,g_t,a_t,r_{i,t},s_{t+1})$ for controller are stored in disjoint spaces for training.

Hierarchical environment includes state, high-level goal, low-level action and reward, which are described specifically below.

\textbf{State:}
State in our hierarchical framework is the same as that in DRL environment, which consists of the channel states between any two nodes in the previous time slot. The expression for the state space can be referred to (\ref{eq4_1}).

\textbf{High-Level Goal:}
In cooperative communication systems, we can intuitively find that relay selection plays a major role.
Therefore, we separate the different action components to make different levels, and extract relay selection as high-level goals for overall planning.
Denote $g_t$ to be the goal in higher level, we then have
\begin{equation}\label{eq5_1} g_t \in \mathcal{G} \triangleq [\boldsymbol g^R(t)]=[g_1^R(t),g_2^R(t),\dots,g_K^R(t)]. \end{equation}

In fact, the goal selection in high-level is similar to relay selection action in the previous DRL method. Therefore, $\boldsymbol g^R(t)$ should meet the same constraint that $\sum_{k=1}^K g_k^R(t)=1,\ g_k^R(t) \in \{0,1\}$.

\textbf{Low-Level Action}:
By decomposing relay selection and power allocation into different levels, we can further reduce the action space. Then the low-level action space only has one variable $\boldsymbol a^{P_s}(t)$, which satisfies $C_2$ and $C_3$ in (\ref{eq4_4_problem}).

\textbf{Reward}:
Note that the higher level and the lower level are working in different time scales. Meta-controller first proposes a temporarily fixed goal for the lower level, and then controller performs actions in a period of time according to both system state and high-level goal and receive feedbacks from the environment.
Therefore, we can denote the internal reward for low-level controller as
\begin{equation}\label{eq5_2} r_{i,t}=1-f\big(\boldsymbol g^{R}(t), \boldsymbol a^{P_s}(t);\boldsymbol h(t)\big). \end{equation}

On the other hand, we use communication success rate of a given relay over a period of time $n$ to measure the quality of current relay selection. Therefore, the external reward for high-level meta-controller can be represented as
\begin{equation}\label{eq5_3} r_{e,t}=\frac{1}{n}\sum\limits_{t=1}^{n} r_{i,t}, \end{equation}
which the agent aims to maximize its expectation.

\newcounter{TempEqCnt} 

\subsection{Hierarchical Learning Policy}\label{subsect_HRLB}
For meta-controller in higher level, we use gradient bandit method to learn goal-policy for dynamically proposing goals according to a given system state.
Recall that, we have $K$ relays to choose from.
Therefore, we first establish the following probability distribution.
\begin{equation}\label{eq5_4} \pi^h_t(R_i) \triangleq Pr\{g_t=R_i\} \triangleq \frac{e^{M_t(R_i)}}{\sum_{b=1}^{K} e^{M_t(R_b)}}, \end{equation}
where $\pi^h$ is the high-level policy, and $\pi^h_t(R_i)$ denotes the probability that relay $R_i$ is selected as the goal in time slot $t$. $M_t(R_b)$ denotes the preference value for choosing relay $R_b$, which will be updated every $n$ steps.

Then, we employ stochastic gradient descent to update the preference values.
\begin{equation}\label{eqappendix_1}
M_{t+n}(R_i) \triangleq M_t(R_i) + \zeta\frac{\partial\mathbb{E}[r_{e,t}]}{\partial{M_t(R_i)}},
\end{equation}
where $\zeta>0$ denotes learning step size, and the expectation of $r_{e,t}$ can be equally calculated by $\sum_{b=1}^K\pi^h_t(R_b)Q^{\ast}_h(R_b)$.
Replace the expectation form in equation (\ref{eqappendix_1}), and we then have
\begin{equation}\label{eqappendix_3} \begin{aligned}
\frac{\partial\mathbb{E}[r_{e,t}]}{\partial{M_t(R_i)}} &= \frac{\partial}{\partial{M_t(R_i)}}\left[\sum\limits_{b=1}^K\pi^h_t(R_b)Q^{\ast}_h(R_b)\right] \\
&= \sum\limits_{b=1}^K Q^{\ast}_h(R_b)\frac{\partial\pi^h_t(R_b)}{\partial{M_t(R_i)}} \\
&= \sum\limits_{b=1}^K (Q^{\ast}_h(R_b)-\bar{r}_{e,t})\frac{\partial\pi^h_t(R_b)}{\partial{M_t(R_i)}} \\
&= \sum\limits_{b=1}^K \pi^h_t(R_b)\frac{(Q^{\ast}_h(R_b)-\bar{r}_{e,t})}{\pi^h_t(R_b)}\frac{\partial\pi^h_t(R_b)}{\partial{M_t(R_i)}}, \\
\end{aligned}\end{equation}
where the newly introduced scalar $\bar{r}_{e,t}$ is independent of $b$. It denotes the average of all external rewards, \textit{i.e.}, the average success rate of our communication system. Further, the partial derivative part can be further written as
\begin{equation}\label{eqappendix_4} \begin{aligned}
\frac{\partial\pi^h_t(R_b)}{\partial{M_t(R_i)}} &= \frac{\partial}{\partial{M_t(R_i)}}\left[\frac{e^{M_t(R_i)}}{\sum_{b=1}^{K} e^{M_t(R_b)}}\right] \\
&= \frac{\mathbbm{1}_{i=b}e^{M_t(R_i)}\sum_{b=1}^{K} e^{M_t(R_b)}-e^{M_t(R_i)}e^{M_t(R_b)}}{\Big(\sum_{b=1}^{K} e^{M_t(R_b)}\Big)^2} \\
&= \mathbbm{1}_{i=b}\pi^h_t(R_b)-\pi^h_t(R_b)\pi^h_t(R_i).
\end{aligned}\end{equation}

Note that, $\mathbb{E}[r_{e,t}|g_t]=Q^{\ast}_h(g_t)$ and $r_{e,t}$ is independent of the other variables.
Then we can derive the following equation by employing the form of expectation.
\begin{equation}\label{eqappendix_5} \begin{aligned}
\frac{\partial\mathbb{E}[r_{e,t}]}{\partial{M_t(R_i)}} &= \sum\limits_{b=1}^K \pi^h_t(R_b)\big(Q^{\ast}_h(R_b)-\bar{r}_{e,t}\big)\big(\mathbbm{1}_{i=b}-\pi^h_t(R_i)\big) \\
&= \mathbb{E}\Big[(r_{e,t}-\bar{r}_{e,t})\big(\mathbbm{1}_{i=b}-\pi^h_t(R_i)\big) \Big].
\end{aligned}\end{equation}

In the training process, sampling is conducted every $n$ time steps, and the gradient in (\ref{eqappendix_1}) is replaced by the expectation value of the single sample. Therefore, we can finally obtain the following update expression of preference value.
\begin{equation}\label{eqappendix_6}
M_{t+n}(R_i) \triangleq M_t(R_i) + \zeta(r_{e,t}-\bar{r}_{e,t})\big(\mathbbm{1}_{i=b}-\pi^h_t(R_i)\big).
\end{equation}

For controller in lower level, it learns action-policy for selecting actions according to both state and goal, which aims to maximize the long-term discounted expected internal reward.

In order to reflect the difference in values of the different actions, we perform some changes to the architecture of traditional deep Q network.
Inspired by \cite{Duel}, we further employ a dueling network, which can enhance the stability of DRL algorithm by ignoring subtle changes in the environment and focusing on key states.

\begin{figure}[ht]
    \centering
    \includegraphics[scale=1.0]{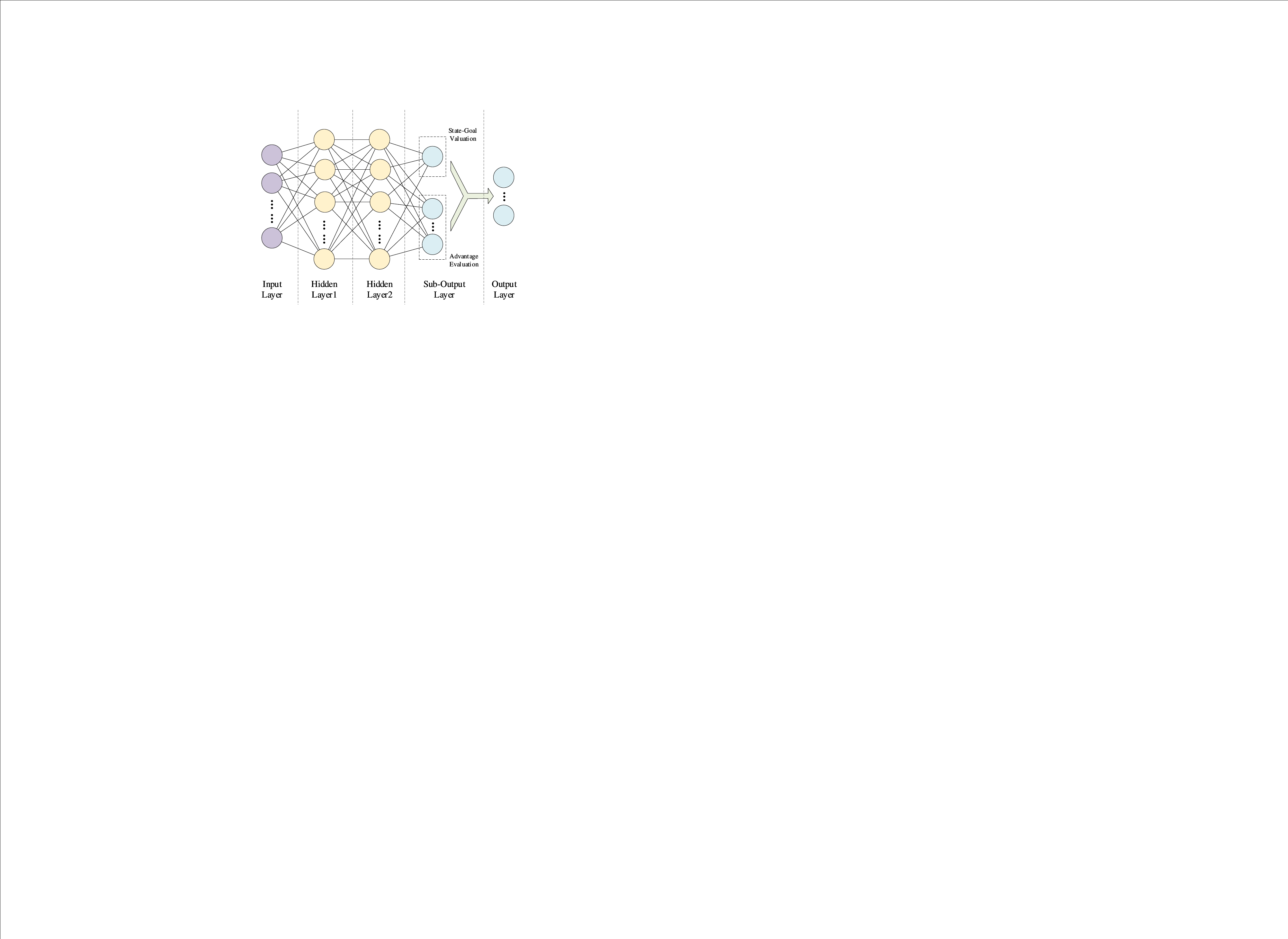}
    \caption{Dueling network in controller for low-level learning.}
    \label{DuelingDNN}
  \end{figure}

Schematic illustration of dueling architecture is shown in Fig. \ref{DuelingDNN}.
The input layer and hidden layers are the same as that in traditional deep Q network.
The key difference is that, there is a sub-output layer in our dueling network, where the traditional Q function output is separated into a state-goal valuation function $V(s_t,g_t)$ and an advantage evaluation function $A(s_t,g_t,a_t)$.
In state-goal valuation part, there is only one neuron which represents the assessment of the current state and goal.
In advantage evaluation part, the number of neurons is equal to that in output layer, representing advantage of choosing each optional action.

Consider the fact that $\mathbb{E}_{a_t\sim\pi_l}[Q(s_t,g_t,a_t)]=V(s_t,g_t)$, we have $\mathbb{E}_{a_t\sim\pi_l}[A(s_t,g_t,a_t)]=0$. Therefore, to meet this property, we can rewrite the advantage part as
\begin{equation}\label{eq_5_5_adv}
A(s_t,g_t,a_t) \triangleq A(s_t,g_t,a_t;\theta,w_2)-\frac{1}{|A|}\sum\limits_{a^{\prime}}A(s_t,a^{\prime};\theta,w_2),
\end{equation}
and thus have the following expression of Q function by combining the two parts in sub-output layer.
\begin{equation}\label{eq_5_5} \begin{aligned}
Q_l(s_t,g_t,a_t) \triangleq V(s_t,g_t;\theta,w_1)+ A(s_t,g_t,a_t;\theta,w_2),
\end{aligned}\end{equation}
where $\theta$ denotes parameters in common part of DNN (\textit{i.e.}, three columns on the left in Fig. \ref{DuelingDNN}), and $w_1$ and $w_2$ are parameters in separated fully connected  sub-output layer for valuation function and advantage function, respectively. Note that, the output of our dueling network is still the same as that of the traditional network, that is, the estimated expected return for each action $a \in \mathcal{A}$ under the current state $s_t$ and goal $g_t$.

By employing temporal difference method, the optimal value of Q function can be written as
\begin{equation}\label{eq5_6} \begin{aligned}
Q_l^{\ast}(s_t,g_t,a_t;\theta,w_1,w_2)
&=\max\limits_{\pi^l_t}\mathbb{E}\left[\sum\limits_{j=t}^{\infty}\gamma_l^{j-t}r_{i,j}\right] \\
&=\max\limits_{\pi^l_t}\mathbb{E}\Big[r_{i,j} + \gamma_l \max\limits_{a_{t+1}}Q_l^{\ast}(s_{t+1},g_t,a_{t+1};\theta,w_1,w_2)\Big],
\end{aligned}\end{equation}
where 
$\pi_t^l$ denotes low-level policy for power allocation in time slot $t$. Note that meta-controller and controller work on different timescales. The controller operates at each time step, while the meta-controller operates on a longer timescale of $n$ time steps.

When training, a batch of memories are sampled from experience replay buffer $B_l$. We calculate a set of loss functions, and derive a batch of gradients as
\begin{equation}\label{eq5_7} \begin{aligned}
\nabla_{\theta_l}L_l(\theta_l)=\mathbb{E}\Big[\big(r_{i,t}+\gamma_l&\max\limits_{a_{t+1}}Q_l(s_{t+1},g_t,a_{t+1};\theta_l^{\prime},w_1,w_2) \\
&-Q_l(s_t,g_t,a_t;\theta_l,w_1,w_2)\big)\nabla_{\theta_l}Q_l(s_t,g_t,a_t;\theta_l,w_1,w_2)\Big].
\end{aligned}\end{equation}
Then the RMSProp optimizer in (\ref{eq4.5_9}) is employed to update network parameters.
Please refer to Algorithm \ref{algo1} for detailed procedure of the hierarchical algorithm.
\begin{algorithm}[htb]
    \caption{HRL Based Relay Selection and Power Allocation}
    \label{algo1}
    \begin{algorithmic}[1]
        \STATE Initialize experience replay buffer: $\mathcal{B}_h$ for higher level, and $\mathcal{B}_l$ for lower level.
        \STATE Initialize deep neural network parameters $\theta_l$ for low-level Q-network.
        \STATE Initialize exploration probability $\epsilon=1$ and anneal factor $\sigma$ for controller.
        \FOR{episode $u=1,2,\dots,u_{max}$}
            \STATE Initialize the environment, obtain initial state $s_0$.
            \STATE Choose goal $g_u$ according to policy $\pi^h$.
            \FOR{time slot $t=1,2,\dots,t_{max}$}
                        \STATE Set goal $g_t=g$ for current time slot.
                        \STATE Choose action $a_t$ using epsilon-greedy method with parameter $\epsilon_{l,g_t}$.
                        \STATE Execute action $a_t$, receive internal reward $r_{i,t}$ from the environment and observe next state $s_{t+1}$.
                        \STATE Collect and save the tuple $(s_t,g_t,a_t,r_{i,t},s_{t+1})$ in $B_l$.
                        \STATE Randomly choose a batch of index, and sample transitions from $B_l$.
                        \STATE Perform gradient descent and update low-level Q-network according to (\ref{eq5_7}).
                        \STATE Update current state.
                \STATE Update internal exploration probability $\epsilon\leftarrow\epsilon-\sigma$.
            \ENDFOR
            \STATE Calculate external reward $r_{e}$ according to (\ref{eq5_3}).
            \STATE Collect and save the tuple $(s_0,g_u,r_{e},s_{t_{max}})$ in $B_h$.
            \STATE Read data from $B_h$, update preference values according to (\ref{eqappendix_6}).
            \STATE Update probability distribution according to (\ref{eq5_4}).
        \ENDFOR
    \end{algorithmic}
\end{algorithm}

\section{Evaluation}\label{sect Result}
In this section, we first introduce the setup of simulation environment. We then carry out experiments to evaluate the proposed algorithms.

\subsection{Experiment Setup}\label{subsect_setup}
Similar to \cite{ref1}, in the two-hop cooperative relay network, channel vectors between any two nodes in each time slot are calculated according to formula (\ref{eq4_2}), where the correlation coefficient $\rho$ is set to be 0.95.
The maximum total power for transmission $P_{max}$, which is the sum of power for source and relay, is limited to 4W.
On the other hand, we set the outage threshold as $\lambda=2.0$ during training process, which means an outage will occur when the MI is lower than 2.0.

To implement our proposed framework, learning step size $\zeta$ in high-level meta-controller is set to be 0.1. Elements in preference value vector are initialized to be 0, and the vector will be updated after each inner loop is completed.

In low-level controller, we use two separate dueling deep Q networks, which share the same structure.
Both dueling deep Q networks include two hidden layers, each of which has 50 neurons, and we employ ReLU function for all hidden layers as activation function.
The number of neurons in input layer is equal to the sum of numbers of states and goals, and the number of neurons in output layer corresponds to the dimensions of low-level action.

Note that, deep Q network requires the action space to be discrete, so we have discretized the power allocation in the environment and set $L$ different power-levels for the agent to choose from. For comparison, we use the following methods as baseline in our experiments.

\textbf{Random Selection}:
For each time slot, the agent randomly selects a relay to perform cooperative communication with random transmission power.

\textbf{DRL Based Approach}:
Our DRL method for minimizing outage probability is proposed in Section \ref{sect DRL}.
We employ traditional DQN framework to make relay selection and power allocation at the same time, and it is now used as one of baseline methods.

\subsection{Numerical Results}\label{subsect_results}
Consider that our DRL method and HRL method both use the structure of deep Q network, therefore we first study the influence of different hyper-parameters on the convergence performance, to obtain the optimal network structure.
Note that, average success rate with different hyper-parameter values is tested 10 times, and mean curves and ranges are then recorded.

\begin{figure}[h]
	\centering
    \includegraphics[scale=0.41]{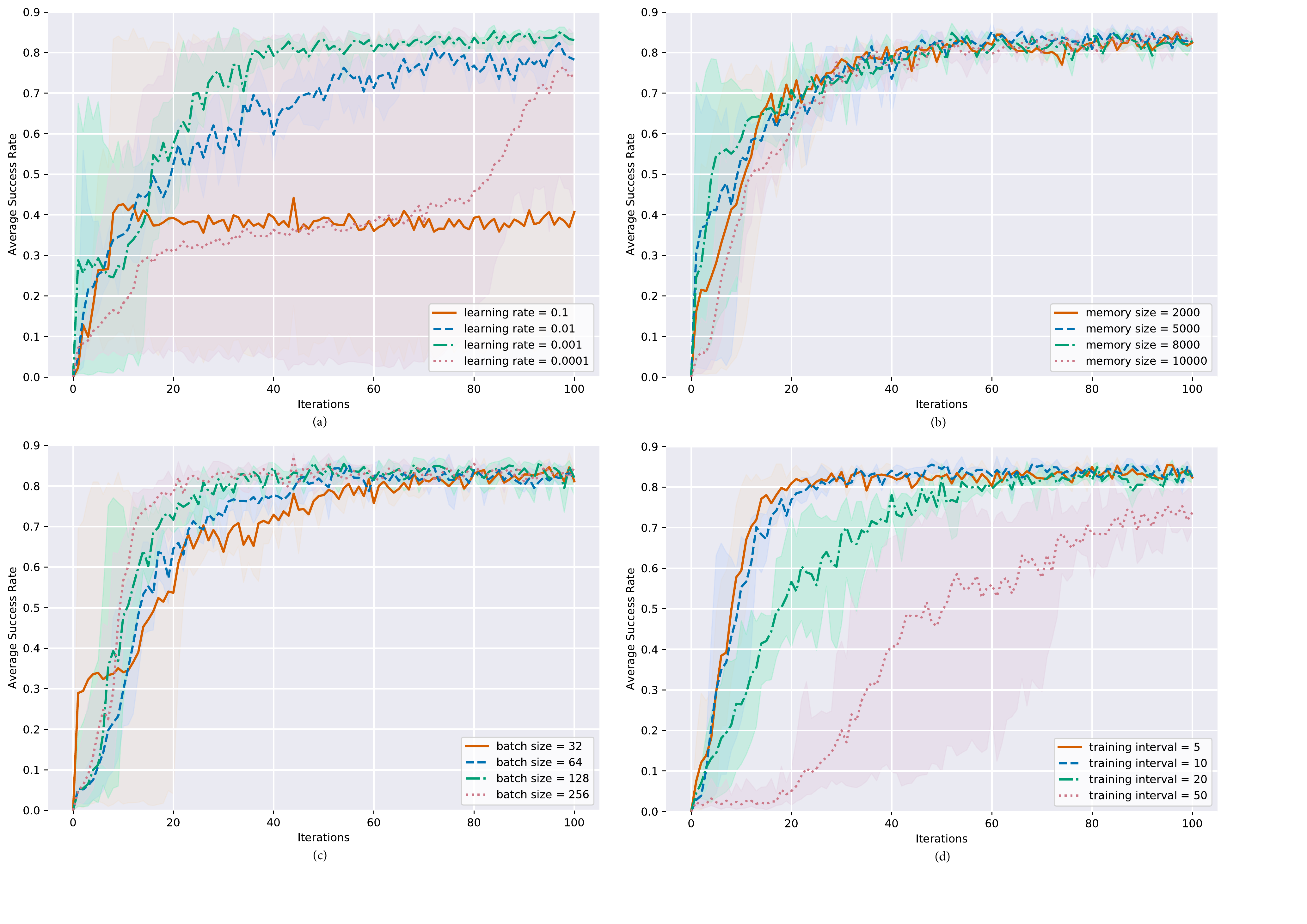}
	\caption{Average success rate under different parameters: (a) learning rate, (b) memory size, (c) batch size, (d) training interval.}
	\label{HyperParameters}
\end{figure}

First of all, the learning rate for updating network parameter should have an appropriate value.
If the learning rate is too small, such as 0.1 (orange solid line), it will lead to local optimum.
If the learning rate is too large, such as 0.0001 (pink dotted line), it will then take much more time to converge.
As shown in Fig. \ref{HyperParameters}(a), we finally set the learning rate as 0.001 for the following simulations.

Experience replay buffer stores experience tuples obtained by the agent.
In Fig. \ref{HyperParameters}(b), we study the effect of replay buffer size, \textit{i.e.}, memory size, on the performance of convergency.
However, unlike Fig. \ref{HyperParameters}(a), different memory sizes have little influence on the final value that average success rate converges to.
Therefore, we directly select the replay buffer size as 8000.

When training, a batch of data is sampled from the experience buffer to improve the DNN.
In Fig. \ref{HyperParameters}(c), we further fix the memory size, and study the effects of different batch sizes during training on the convergence performance.
It can be found that training with a small batch size cannot take advantage of all data stored in experience buffer, and converges slowly.
While training with a large batch size, such as 256 (pink dotted line), has the fastest convergence speed, although it will consume much more time during training process.

Finally, we investigate the convergence performance under different training intervals, as shown in Fig. \ref{HyperParameters}(d).
Theoretically, the shorter the training interval, the faster the convergence speed.
However, shorter training interval also means more training times, which will result in a certain waste of computing resources.
On the other hand, we find that the final convergence values with time intervals of 5 and 10 are very close.
Considering the above reasons, we finally set training interval as 10.

Set the above hyper-parameters to optimal values and apply them to all deep Q networks, then we carry out the following experiments.

\begin{figure}[ht]
	\centering
    \includegraphics[scale=0.6]{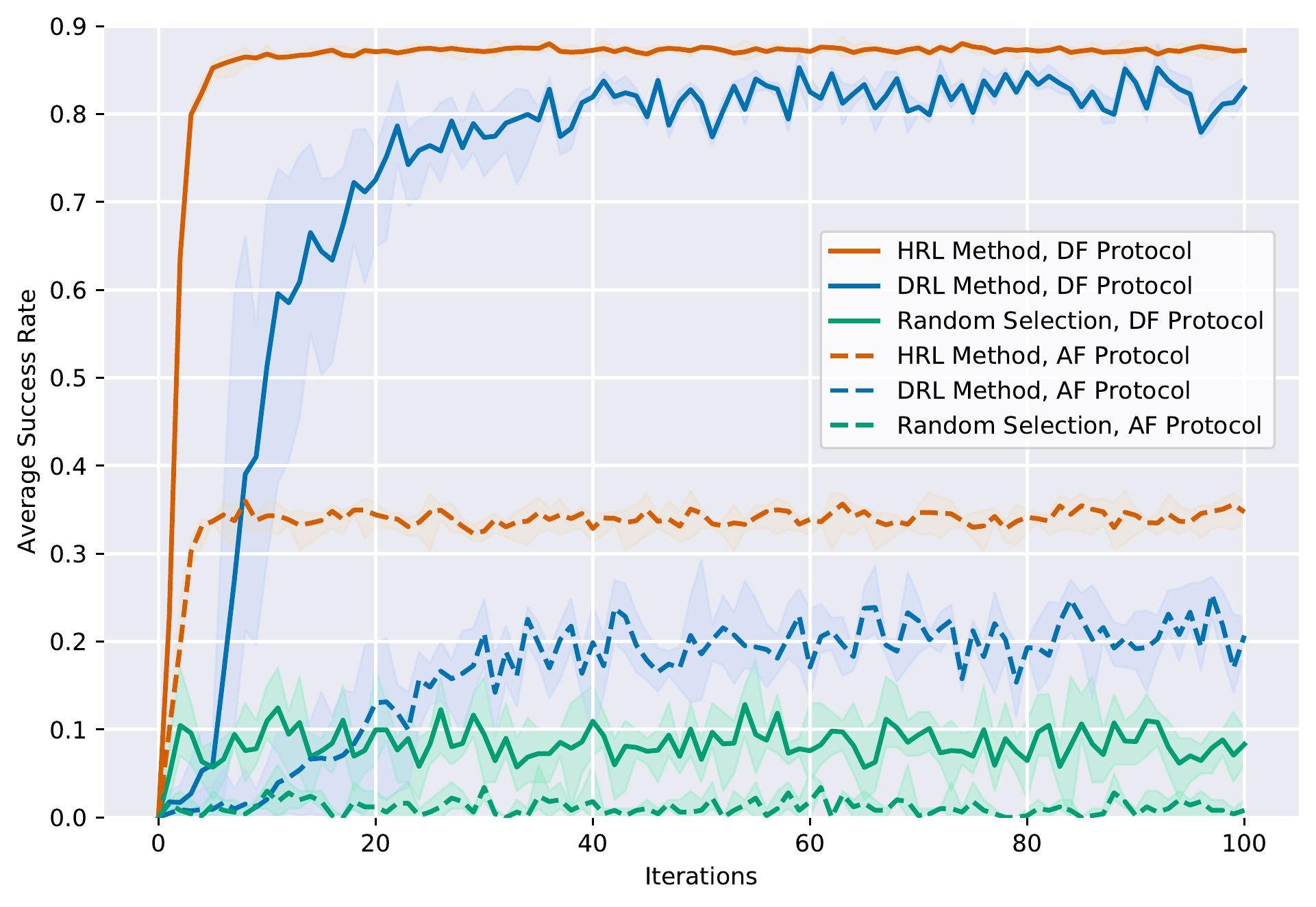}
	\caption{Average success rate using different protocols.}
	\label{TrainReward}
\end{figure}

In training process, we evaluate the performance of different methods in AF environment and DF environment, with relay number $K=10$ and power level number $L=10$. The result is depicted in Fig. \ref{TrainReward}.

It can be observed that when using the method of random selection, the performance is always very poor.
On the other hand, both DRL method and our hierarchical algorithm can be effectively trained, and their average success rate curves eventually converge to a stable value with slight changes.

However, our HRL method can achieve a lower outage probability. Take employing DF protocol as an example, with DRL method, the average success rate is only about 0.82, which means the outage probability of communication system is about $18\%$.
When employing our hierarchical method, the average success rate is closer to 0.9, with an improvement of about $5\%$.
On the other hand, our HRL method has an obvious faster learning speed, which converges after about 10 iterations, while DRL method needs about 40 iterations to reach the convergence value.

When comparing results under with different protocols, Fig. \ref{TrainReward} shows that there is little difference in convergence speed of the same method, but the fluctuation of convergence value using DRL method is larger.
We also observe that the average success rate using AF protocol is generally lower than that using DF protocol, which leads to less successful experiences for agents to learn from.
The performance of DRL method is obviously affected by this factor, but our HRL method can still converge to a value with slighter changes.
What's more, in the case of using AF protocol, the average communication success rate obtained by our HRL method is approximately $15\%$ higher than that obtained by DRL method.
Compared with DF protocol, the performance gap between the two methods under AF protocol is larger.
In all, our HRL agent can learn a better strategy faster for dynamic relay selection and power allocation, in both AF communication environment and DF communication environment.

\begin{figure}[ht]
	\centering
    \includegraphics[scale=0.6]{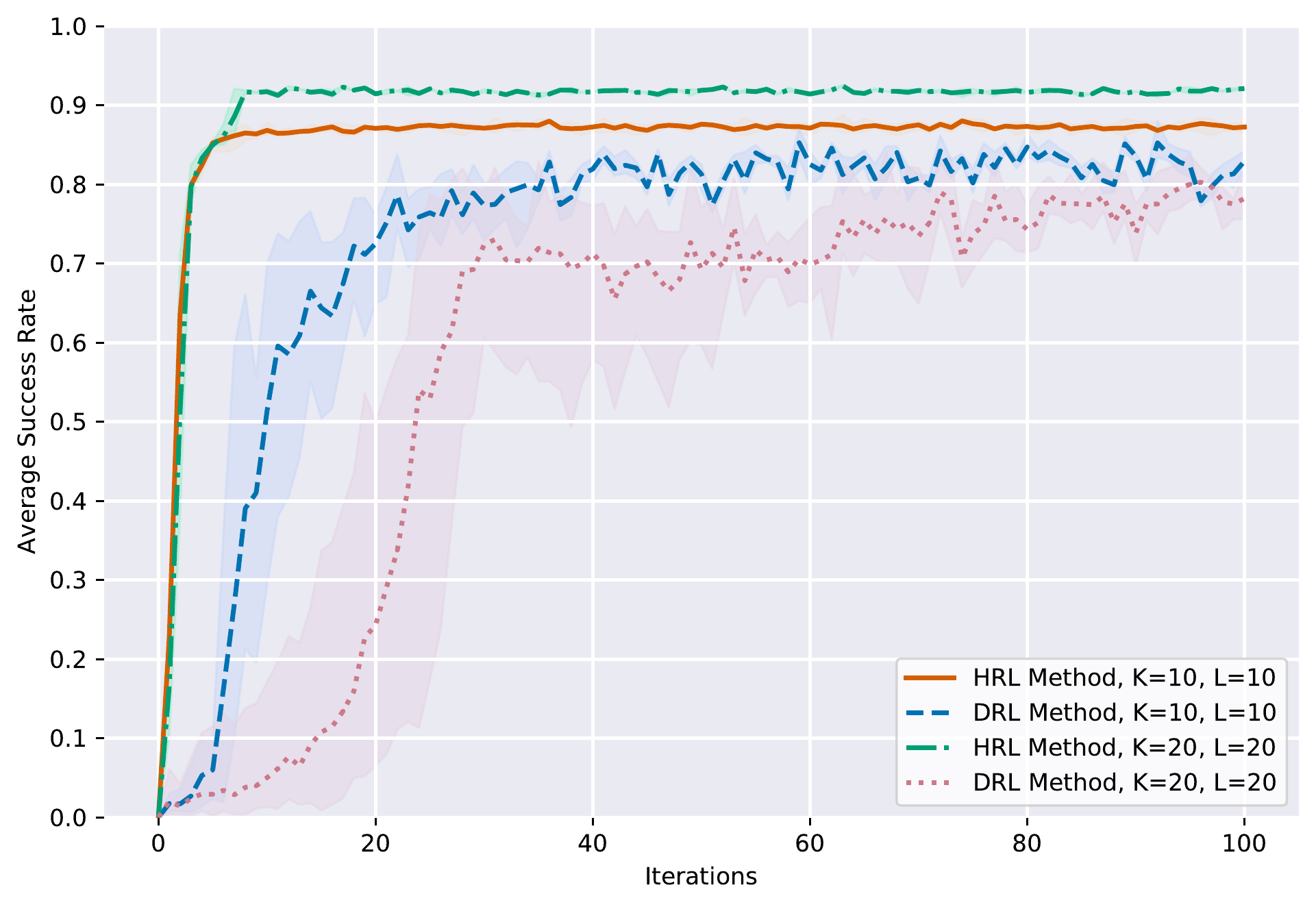}	
    \caption{Outage probability under different relay number $K$ and power level number $L$.}
	\label{TestKL}
\end{figure}

Then, we evaluate the performance of our proposed hierarchical method and DRL method under different search space scales.
We conduct this experiment in DF communication environment. As shown in Fig. \ref{TestKL}, we study two scenarios where the number of relays $K$ and power levels $L$ are both set to be 10 or 20.

As $K$ and $L$ increased, there is a small increase in the average success rate of our HRL method, which increases from 0.87 to 0.91.
With optional power level $L$ increased, transmission power can be allocated more efficiently at source and relay node, resulting in an improved communication success rate.
In addition, Fig. \ref{TestKL} also vividly shows that the result curve obtained by our HRL method is much smoother than that obtained by DRL method.

One the other hand, DRL method performs worse under larger $K$ and $L$, where the average success rate drops by about $5\%$ after convergence.
In a larger search space, it becomes more difficult for the agent to select appropriate relay and power simultaneously.
As a result, there are fewer successful explorations, which leads to a problem of sparse reward.
Therefore, it takes more training iterations for DRL agent to converge, and the fluctuation becomes larger.
It is worth noting that the number of successful explorations is also small when using the AF protocol (in Fig. \ref{TrainReward}), but this situation is actually different from that shows in Fig. \ref{TestKL}.
In the previous experiment, even if the optimal action is taken, there is still a high probability that the communication will fail due to the uncertainty of the channel.
However, in this experiment, the optimal action with good return exists, but the agent may not be able to find this action policy as the search space is too large.
Therefore, the problem of sparse reward usually refers to the latter situation.

Traditional DRL methods usually perform poorly in the environment with sparse rewards, due to the lack of positive experience to learn from.
However, by making different hierarchies, our method can reduce the complexity of search space, which ensures the efficiency of exploration and learning.
Therefore, when employing our proposed hierarchical method, we can still obtain a more stable behavior policy for relay selection and power allocation.

\begin{figure}[ht]
	\centering
	\includegraphics[scale=0.75]{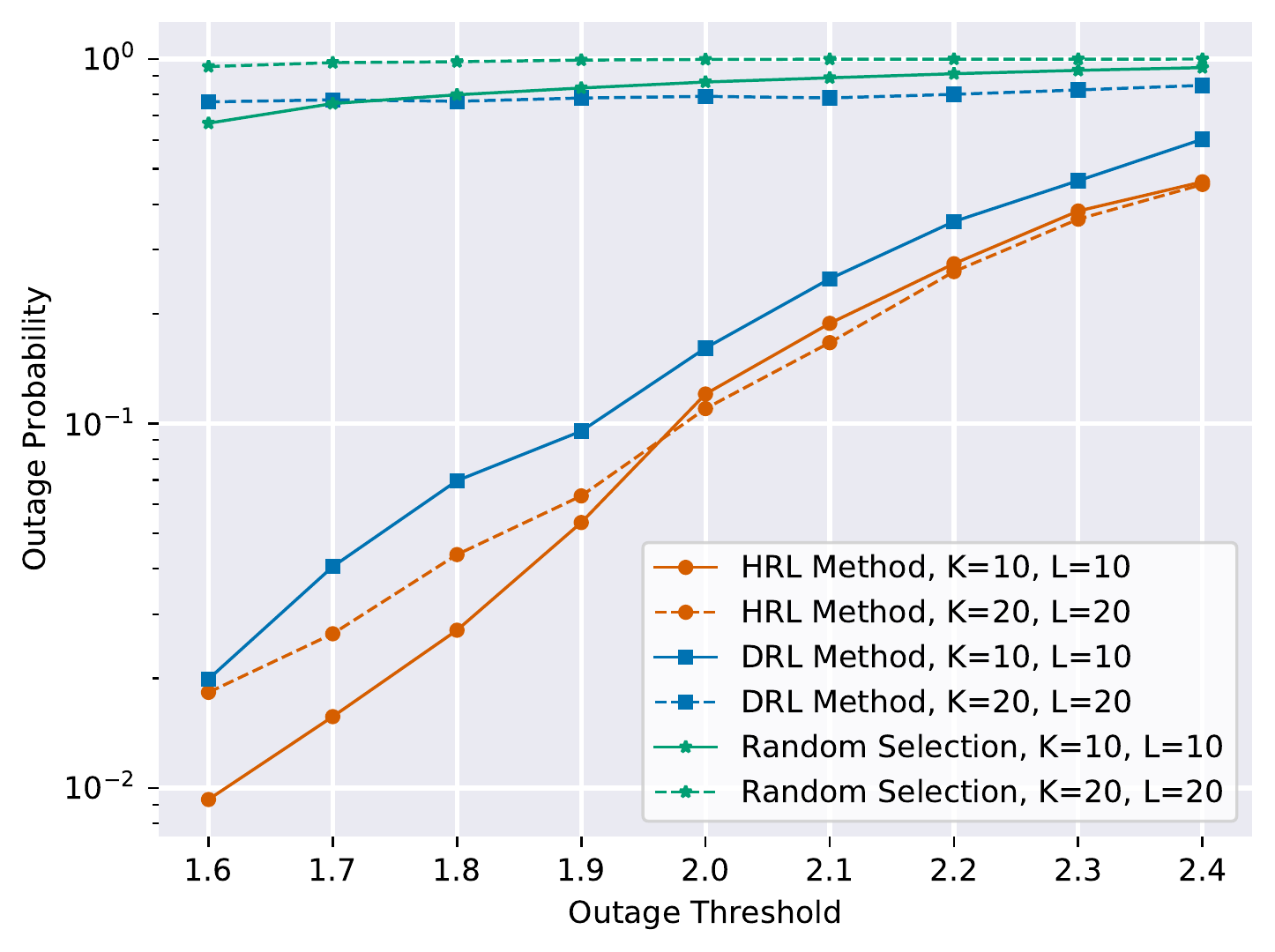}
    \caption{Outage probability under different outage threshold $\lambda$.}
	\label{TestReward}
\end{figure}

After 100 iterations of training, we obtain dynamic relay selection and power allocation policies by applying both HRL method and DRL based method.
To further evaluate the robustness of different methods, we evaluate the performance by using these well-trained policies under different outage thresholds, and the result is depicted in Fig. \ref{TestReward}.

This experiment is conducted in DF environment, and the communication outage threshold ranges from 1.6 to 2.4 in testing process.
The only difference between testing and training is that, the parameters of all networks in testing process are fixed, which means DNN is only used to provide a best action rather than executing further learning.

As we can see from Fig. \ref{TestReward}, both HRL policy and DRL policy trained in a smaller search space can be applied to other situations.
However, DRL policy trained in a larger search space performs poorly when testing, while we can still obtain proper actions in different environments by following our HRL policy.

Take $\lambda=1.8$ as an example, the outage probability under different policies is different.
In terms of the small search space, the outage probability using HRL policy is lower than 0.03, and that using DRL policy and random selection are about 0.07 and 0.8.
In terms of the large search space, outage probability using HRL policy is about 0.04, and that using DRL policy and random selection are 0.8 and nearly 1.0.
It is obvious that our HRL method is more robust and can greatly reduce the outage probability, which means that HRL agent can perform better relay selection and adjust power allocation more reasonably according to the current state after training.

\section{Conclusion and Future Works}\label{sect Conclusion}
In this paper, we propose an HRL method to dynamically select relay and allocate transmission power in a two-hop cooperative communication model, in order to minimize outage probability under a total transmission power constraint.
Unlike traditional studies, our method does not require any assumptions about channel distribution, but relies on the interaction between the agent and the communication environment.
Compared with existing RL-based methods, we propose an outage-based reward function.
Our reward function uses only binary reward that indicates the result of communication, while other RL-based methods require concrete representations of feedback information, such as instantaneous SNR or MI.
We further design an HRL framework by decomposing relay selection and power allocation into two sub-tasks, which reduces the search space.
Simulation results show that our HRL method can reduce outage probability by $5\%$, and reach convergence 30 iterations earlier than DRL methods in both AF and DF communication environment.
In addition, the hierarchical method can effectively solves the problem of sparse reward, while other methods can hardly deal with it.

Our HRL method provides a novel way for the research of resource allocation and optimization in the field of communication. However, the total transmission power is discretized into enumerable power levels in our framework, which can be improved further. In future works, we would like to explore new methods applicable to continuous action space.



\bibliographystyle{IEEEtran}
\bibliography{paper_ref}

\begin{thebibliography}{10}
\providecommand{\url}[1]{#1}
\csname url@samestyle\endcsname
\providecommand{\newblock}{\relax}
\providecommand{\bibinfo}[2]{#2}
\providecommand{\BIBentrySTDinterwordspacing}{\spaceskip=0pt\relax}
\providecommand{\BIBentryALTinterwordstretchfactor}{4}
\providecommand{\BIBentryALTinterwordspacing}{\spaceskip=\fontdimen2\font plus
\BIBentryALTinterwordstretchfactor\fontdimen3\font minus
  \fontdimen4\font\relax}
\providecommand{\BIBforeignlanguage}[2]{{%
\expandafter\ifx\csname l@#1\endcsname\relax
\typeout{** WARNING: IEEEtran.bst: No hyphenation pattern has been}%
\typeout{** loaded for the language `#1'. Using the pattern for}%
\typeout{** the default language instead.}%
\else
\language=\csname l@#1\endcsname
\fi
#2}}
\providecommand{\BIBdecl}{\relax}
\BIBdecl

\bibitem{8275026}
F.~{Zhong}, X.~{Xia}, H.~{Li}, and Y.~{Chen}, ``Distributed linear
  convolutional space-time coding for two-hop full-duplex relay 2x2x2
  cooperative communication networks,'' \emph{IEEE Transactions on Wireless
  Communications}, vol.~17, no.~5, pp. 2857--2868, May 2018.

\bibitem{7886242}
C.~{Wang}, T.~{Cho}, T.~{Tsai}, and M.~{Jan}, ``A cooperative multihop
  transmission scheme for two-way amplify-and-forward relay networks,''
  \emph{IEEE Transactions on Vehicular Technology}, vol.~66, no.~9, pp.
  8569--8574, Sept. 2017.

\bibitem{Liu2012:Energy}
Y.~{Liu}, E.~{Liu}, and R.~{Wang}, ``Energy efficiency analysis of intelligent
  reflecting surface system with hardware impairments,'' in \emph{2020 IEEE
  Global Communications Conference: Wireless Communications (Globecom2020 WC)},
  Taipei, Taiwan, December 2020.

\bibitem{8435942}
Y.~{Shi}, A.~{Konar}, N.~D. {Sidiropoulos}, X.~{Mao}, and Y.~{Liu}, ``Learning
  to beamform for minimum outage,'' \emph{IEEE Transactions on Signal
  Processing}, vol.~66, no.~19, pp. 5180--5193, Oct. 2018.

\bibitem{2006.00664}
Y.~{Liu}, E.~{Liu}, R.~{Wang}, and Y.~{Geng}, ``Beamforming designs and
  performance evaluations for intelligent reflecting surface enhanced wireless
  communication system with hardware impairments,'' \emph{arXiv preprint
  arXiv:2006.00664}, 2020.

\bibitem{7438886}
J.~{Jedrzejczak}, G.~J. {Anders}, M.~{Fotuhi-Firuzabad}, H.~{Farzin}, and
  F.~{Aminifar}, ``Reliability assessment of protective relays in
  harmonic-polluted power systems,'' \emph{IEEE Transactions on Power
  Delivery}, vol.~32, no.~1, pp. 556--564, Feb. 2017.

\bibitem{7778750}
S.~N. {Islam}, M.~A. {Mahmud}, and A.~M.~T. {Oo}, ``Relay aided smart meter to
  smart meter communication in a microgrid,'' in \emph{2016 IEEE International
  Conference on Smart Grid Communications (SmartGridComm)}, Sydney, NSW,
  Australia, Nov. 2016, pp. 128--133.

\bibitem{7105886}
P.~{Das} and N.~B. {Mehta}, ``Direct link-aware optimal relay selection and a
  low feedback variant for underlay {CR},'' \emph{IEEE Transactions on
  Communications}, vol.~63, no.~6, pp. 2044--2055, Jun. 2015.

\bibitem{1603719}
A.~{Bletsas}, A.~{Khisti}, D.~P. {Reed}, and A.~{Lippman}, ``A simple
  cooperative diversity method based on network path selection,'' \emph{IEEE
  Journal on Selected Areas in Communications}, vol.~24, no.~3, pp. 659--672,
  Mar. 2006.

\bibitem{7313006}
C.~{Wang} and J.~{Chen}, ``Power allocation and relay selection for af
  cooperative relay systems with imperfect channel estimation,'' \emph{IEEE
  Transactions on Vehicular Technology}, vol.~65, no.~9, pp. 7809--7813, Sept.
  2016.

\bibitem{5722051}
O.~{Amin}, S.~S. {Ikki}, and M.~{Uysal}, ``On the performance analysis of
  multirelay cooperative diversity systems with channel estimation errors,''
  \emph{IEEE Transactions on Vehicular Technology}, vol.~60, no.~5, pp.
  2050--2059, Jun. 2011.

\bibitem{6108302}
M.~{Seyfi}, S.~{Muhaidat}, and J.~{Liang}, ``Amplify-and-forward selection
  cooperation over {Rayleigh} fading channels with imperfect {CSI},''
  \emph{IEEE Transactions on Wireless Communications}, vol.~11, no.~1, pp.
  199--209, Jan. 2012.

\bibitem{5618893}
F.~S. {Tabataba}, P.~{Sadeghi}, and M.~R. {Pakravan}, ``Outage probability and
  power allocation of amplify and forward relaying with channel estimation
  errors,'' \emph{IEEE Transactions on Wireless Communications}, vol.~10,
  no.~1, pp. 124--134, Jan. 2011.

\bibitem{6954557}
F.~{Shams}, G.~{Bacci}, and M.~{Luise}, ``Energy-efficient power control for
  multiple-relay cooperative networks using {Q}-learning,'' \emph{IEEE
  Transactions on Wireless Communications}, vol.~14, no.~3, pp. 1567--1580,
  Mar. 2015.

\bibitem{9072416}
X.~{Wang}, T.~{Jin}, L.~{Hu}, and Z.~{Qian}, ``Energy-efficient power
  allocation and {Q}-learning-based relay selection for relay-aided {D2D}
  communication,'' \emph{IEEE Transactions on Vehicular Technology}, vol.~69,
  no.~6, pp. 6452--6462, Jun. 2020.

\bibitem{8750861}
Y.~{Su}, X.~{Lu}, Y.~{Zhao}, L.~{Huang}, and X.~{Du}, ``Cooperative
  communications with relay selection based on deep reinforcement learning in
  wireless sensor networks,'' \emph{IEEE Sensors Journal}, vol.~19, no.~20, pp.
  9561--9569, Oct. 2019.

\bibitem{9137340}
Y.~{Su}, M.~{LiWang}, Z.~{Gao}, L.~{Huang}, X.~{Du}, and M.~{Guizani},
  ``Optimal cooperative relaying and power control for {IoUT} networks with
  reinforcement learning,'' \emph{IEEE Internet of Things Journal}, pp. 1--1,
  Jul. 2020.

\bibitem{8931561}
Y.~{Hua}, R.~{Li}, Z.~{Zhao}, X.~{Chen}, and H.~{Zhang}, ``{GAN}-powered deep
  distributional reinforcement learning for resource management in network
  slicing,'' \emph{IEEE Journal on Selected Areas in Communications}, vol.~38,
  no.~2, pp. 334--349, Feb. 2020.

\bibitem{8761525}
L.~P. {Qian}, A.~{Feng}, X.~{Feng}, and Y.~{Wu}, ``Deep {RL}-based time
  scheduling and power allocation in {EH} relay communication networks,'' in
  \emph{IEEE International Conference on Communications (ICC)}, Shanghai,
  China, May 2019, pp. 1--7.

\bibitem{8771176}
L.~{Huang}, S.~{Bi}, and Y.~J.~A. {Zhang}, ``Deep reinforcement learning for
  online computation offloading in wireless powered mobile-edge computing
  networks,'' \emph{IEEE Transactions on Mobile Computing}, vol.~19, no.~11,
  pp. 2581--2593, Nov. 2020.

\bibitem{RL:intro2}
R.~S. {Sutton} and A.~G. {Barto}, \emph{Reinforcement learning: An
  introduction}.\hskip 1em plus 0.5em minus 0.4em\relax MIT press, 2018.

\bibitem{5710995}
H.~A. {Suraweera}, T.~A. {Tsiftsis}, G.~K. {Karagiannidis}, and
  A.~{Nallanathan}, ``Effect of feedback delay on amplify-and-forward relay
  networks with beamforming,'' \emph{IEEE Transactions on Vehicular
  Technology}, vol.~60, no.~3, pp. 1265--1271, Mar. 2011.

\bibitem{1427716}
A.~{Ribeiro}, {Xiaodong Cai}, and G.~B. {Giannakis}, ``Symbol error
  probabilities for general cooperative links,'' \emph{IEEE Transactions on
  Wireless Communications}, vol.~4, no.~3, pp. 1264--1273, May 2005.

\bibitem{1350931}
J.~{Boyer}, D.~D. {Falconer}, and H.~{Yanikomeroglu}, ``Multihop diversity in
  wireless relaying channels,'' \emph{IEEE Transactions on Communications},
  vol.~52, no.~10, pp. 1820--1830, Oct. 2004.

\bibitem{4107949}
R.~{Annavajjala}, P.~C. {Cosman}, and L.~B. {Milstein}, ``Statistical channel
  knowledge-based optimum power allocation for relaying protocols in the high
  {SNR} regime,'' \emph{IEEE Journal on Selected Areas in Communications},
  vol.~25, no.~2, pp. 292--305, Feb. 2007.

\bibitem{ref1}
Z.~{Chen} and X.~{Wang}, ``Decentralized computation offloading for multi-user
  mobile edge computing: A deep reinforcement learning approach,'' \emph{arXiv
  preprint arXiv:1812.07394}, 2018.

\bibitem{Mnih2015}
V.~Mnih, K.~Kavukcuoglu, D.~Silver \emph{et~al.}, ``Human-level control through
  deep reinforcement learning,'' \emph{Nature}, vol. 518, no. 7540, pp.
  529--533, Feb. 2015.

\bibitem{IMPALA}
L.~{Espeholt}, H.~{Soyer}, R.~{Munos} \emph{et~al.}, ``{IMPALA}: {Scalable}
  distributed deep-{RL} with importance weighted actor-learner architectures,''
  in \emph{International Conference on Machine Learning (ICML)}, Stockholm,
  Sweden, Jul. 2018, pp. 1407--1416.

\bibitem{A3C}
V.~{Mnih}, A.~P. {Badia}, and M.~{Mirza}, ``Asynchronous methods for deep
  reinforcement learning,'' in \emph{International Conference on Machine
  Learning (ICML)}, New York City, NY, USA, Jun. 2016, pp. 1928--1937.

\bibitem{lecture}
T.~{Tieleman}, G.~{Hinton}, G.~K. {Karagiannidis}, and A.~{Nallanathan},
  ``Lecture 6.5-rmsprop: {Divide} the gradient by a running average of its
  recent magnitude,'' \emph{COURSERA: Neural Networks for Machine Learning},
  vol.~4, no.~2, pp. 26--31, 2012.

\bibitem{HDQN}
T.~D. {Kulkarni}, K.~{Narasimhan}, A.~{Saeedi}, and J.~{Tenenbaun},
  ``Hierarchical deep reinforcement learning: {Integrating} temporal
  abstraction and intrinsic motivation,'' in \emph{Neural Information
  Processing Systems (NIPS)}, Barcelona, Spain, Dec. 2016, pp. 3675--3683.

\bibitem{HIRO}
O.~{Nachum}, S.~{Gu}, H.~{Lee}, and S.~{Levine}, ``Data-efficient hierarchical
  reinforcement learning,'' in \emph{Neural Information Processing Systems
  (NIPS)}, Montreal, Canada, Dec. 2018, pp. 3303--3313.

\bibitem{8629360}
N.~{Dilokthanakul}, C.~{Kaplanis}, N.~{Pawlowski}, and M.~{Shanahan}, ``Feature
  control as intrinsic motivation for hierarchical reinforcement learning,''
  \emph{IEEE Transactions on Neural Networks and Learning Systems}, vol.~30,
  no.~11, pp. 3409--3418, Nov. 2019.

\bibitem{Duel}
Z.~{Wang}, T.~{Schaul} \emph{et~al.}, ``Dueling network architectures for deep
  reinforcement learning,'' in \emph{International Conference on Machine
  Learning (ICML)}, New York City, NY, USA, Jun. 2016, pp. 1995--2003.

\end{thebibliography}

\end{document}